# Goal-oriented Composition of Software Process Patterns

Jürgen Münch
*Fraunhofer Institute for Experimental Software Engineering*
*Sauerwiesen 6, 67661 Kaiserslautern, Germany*
*Juergen.Muench@iese.fraunhofer.de*

**Abstract**

*The development of high-quality software or software-intensive systems requires custom-tailored process models that fit the organizational and project goals as well as the development contexts. These models are a necessary prerequisite for creating project plans that are expected to fulfill business goals. Although project planners require individual process models custom-tailored to their constraints, software or system developing organizations also require generic processes (i.e., reference processes) that capture project-independent knowledge for similar development contexts. The latter is emphazised by assessment approaches (such as CMMI, SPICE) that require explicit process descriptions in order to reach a certain capability or maturity level. Among other concepts such as polymorphism, templates, or generator-based descriptions, software process patterns are used to describe generic process knowledge. Several approaches for describing the architecture of process patterns have already been published (e.g., [7]). However, there is a lack of descriptions on how to compose process patterns for a specific decelopment context in order to gain a custom-tailored process model for a project. This paper focuses on the composition of process patterns in a goal-oriented way. First, the paper describes which information a process pattern should contain so that it can be used for systematic composition. Second, a composition method is sketched. Afterwards, the results of a proof-of-concept evaluation of the method are described. Finally, the paper is summarized and open research questions are sketched.*

## 1. Introduction

Providing generic processes for organizations and deriving custom-tailored software development processes for projects is a challenging task: Typically, contexts and technology are changing often, new state-of-the art knowledge should be considered, conformance to standards often needs to be guaranteed, experience from past projects should be captured, and distributed development (such as offshoring) has a growing impact on software development processes. Besides these aspects, process models need to be designed, described, evolved, and introduced in a way that they are really used in practice. This is a challanging task requiring methodological, technological, and organizational support for software developing organizations.

Two levels of abstraction can be distinguished: On the one hand, generic processes capture project-independent process knowledge that is relevant for certain application domains (e.g., space software) and/or development contexts (e.g., large projects). On the other hand, project-specific process models describe activities for concrete projects. Typically, project-specific models are derived from generic models through tailoring.

Both generic models and project-specific models need to be evolved: Process-specific models are often evolved through refinement or replanning activities during the course of the project. Generic processes can be evolved, for instance, by including feedback from the execution of projects that are in the scope of the generic model.

The tailoring of software process models to project constraints requires an understanding of process variations and knowledge about when to use which variation. Typically, the following logical steps are needed before the customization is performed [11]:

- Possible process alternatives need to be elicited and explicitly described.
- Process alternatives need to be characterized and constraints/rules on their use need to be formulated. This requires a deeper understanding of the appropriateness of the process alternatives for different contexts and their effects in these contexts.
- Before the start of the project, a characterization of the project context and its goal is necessary, describing the information needed to select process alternatives.

The approach described in this article is based on an engineering-style development paradigm. This means that planning is based on explicit experience from past projects (i.e., models) and is done in a goal-oriented way. The performance of development activities should adhere to the plan and appropriate means should be used to control and avoid plan deviations or to do replanning in a systematic way. Finally, experience gained during project execution should be analyzed and captured for future projects. Due to the context-orientation of software development (i.e., development activities are unique for a specific context), models need to be customizable to contexts and their scope of validity should be defined. This means that the applicable contexts and the degree of evaluation must be attached to a model.

Software process patterns can be seen as process model fragments that capture relevant process information for solving typical software or system engineering problems in specific contexts. Patterns are a reuse mechanism for software knowledge. The origins of patterns lie in the building architecture domain [1]. In software engineering, first patterns were used in the area of software design [5]. Based on typical characteristics of these domains, patterns seem to be appropriate for capturing knowledge, at least knowledge about creative activities. This could be seen as a hint that patterns are also an appropriate means for capturing knowledge about software development processes that typically also include highly creative activities. Besides process patterns, patterns are nowadays also used in other software engineering areas such as requirements engineering (e.g., requirements engineering patterns [8]) or organizational structuring (e.g., organizational patterns [3][9]).

Applying software process patterns in the context of project planning requires a set of patterns, a schema for characterizing those patterns, an approach for selecting appropriate patterns, an approach for composing the selected patterns, and, finally, a technique for integrating those patterns on the level of a process modeling notation. This paper focuses on the composition. Approaches for characterization and selection are described, for instance, in [2],[6],[12]. An approach for integrating patterns (or process fragments) on the level of a formal process modeling language is described in [11].

## 2. Process Patterns and Goals

We define a process pattern as a reusable fragment of a process model or a combination of process models that represents an activity or a set of activities, and that is described together with the following information (see Figure 1):

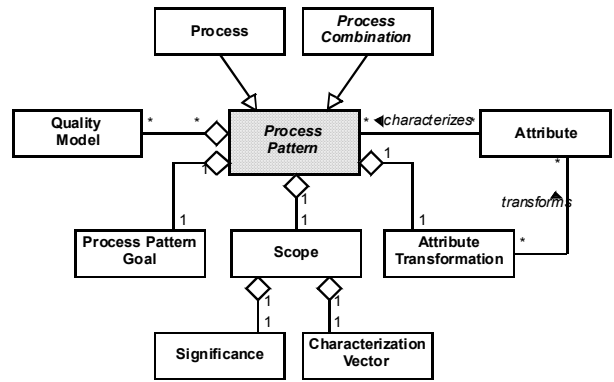

**Figure 1. Process Pattern Architecture.**

- Characterization vector: a set of characterizing attributes C and an assignment of values C(t) at a certain point in time t.
- Significance: information about the degree of validation of the pattern (e.g., information about the number of usages of the pattern in the context defined by the characterization vector).
- Process pattern goal: an atomic goal describes a restriction on the attributes from the characterization vector by using $a < b$, $a <= b$, $a > b$, $a >= b$, $a = b$ or $a \in M$. A complex goal is a logic combination of atomic goals or complex goals using the operators "&" (and), "|" (or), "¬" (negation), "⇒" (implication) and "⇔" (equivalence). Process pattern goals can be used to describe the goal that can be reached by a pattern in a certain context. If the pattern is composed of other patterns, the goal describes the goal that can be reached by the composition. An example goal is ((maximal defect rate < 0.8 %) & (service level = high) & (test-effort = f(design-complexity)))
- Attribute transformation: An attribute transformation describes a change of attributes that is caused by the application of a process pattern. An example is reliability:=$_P$ reliability * 1.15, where p is the process pattern.
- Quality models: models describing cause-effect relations (e.g., a prediction model for test effort based on design complexity) can be associated to a pattern.
- Process combination: Process patterns can be combined from other process patterns using composition operators (see below).

The description of pattern goals, attribute transformations and quality models should ideally be based on empirical evidence. Also, process simulation could be used to get relevant knowledge about the effects of process patterns in certain contexts.

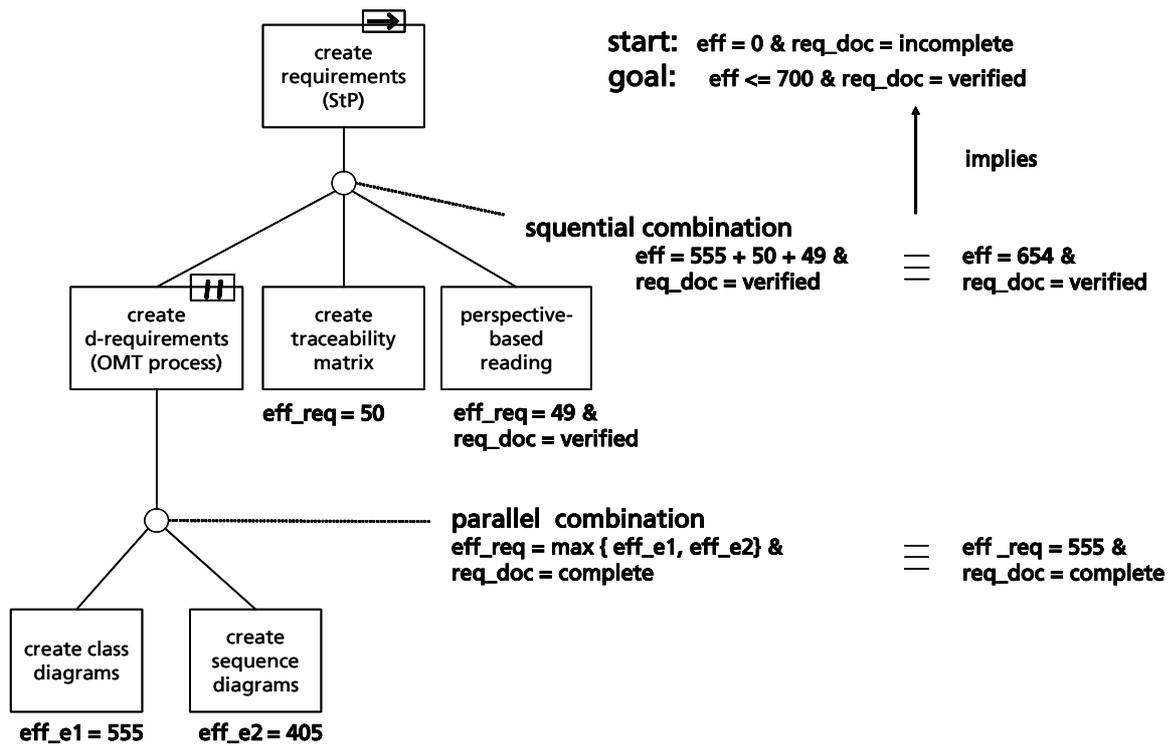

**Figure 2. Goal-oriented Composition of Process Patterns.**

## 3. Process Pattern Composition

The starting point for a goal-oriented process pattern composition are project goals that can be formulized analogous to the process pattern goals. The attributes that can be used in the goal definition are a subset of the characterizing attributes of the process patterns and the attributes describing the project context.

For the composition of process patterns, the following composition operators have been defined: sequential combination, parallel combination, conditional combination, and iterative combination. These operators describe how process patterns can be composed and the implications of the compositions on the attribute mappings. The iterative combination can be defined recursively by using the conditional combination:

while ( $c$ ) $t$; ≡ if ( $c$ ) $t$; while ( $c$ ) $t$;

The overall composition method consists of the following steps:
1. Composition of process patterns by replacement, refinement, and augmentation. Based on an initial process pattern, this step creates a combination of process patterns. How process patterns can be combined is constrained. Such constraints are, for instance, that two process patterns can only be combined if the same tool is used or that a logical control flow sequence for a set of process patterns is necessary because of necessary product flows. Such constraints are described via so-called composition networks that need to be established for application domains (see [10] for further details).
2. Verification of the process pattern combination by using formal semantics. During this step, the question is answered of whether the current process pattern combination fulfills the project goal (or sub-goals).
3. If the verification is successful, the formal process descriptions of the process patterns need to be integrated. The approach described in [11] and [13], which also integrates the product flows between the process patterns, could be used. This step includes the transformation of processes from the pattern level to the process instance level, i.e., the customization of the integrated process patterns to the project environment. If the verification is not successful, a new composition needs to be chosen (i.e., step 1 starts again).

Verification and goal-oriented composition of process patterns can now be defined in the following way: Let g be a project goal and p a process pattern combination. Let [p] denote the attribute transformation that is caused by the process pattern combination p. Let s be the values of all characterizing attributes at the beginning of the project. Then we can define:

1. The verification of a goal is the proof that
   $[p](s) \Rightarrow g$
2. The goal-oriented composition of process patterns is the creation of a process pattern combination p so that $[p](s) \Rightarrow g$

The details of the method are described in [10]. Figure 2 gives an example of the method: Here, the project goal is the development of a verified requirements document with an overall effort (eff) of less than 700 hours. The starting conditions are that no effort has been consumed and the requirements document is incomplete. Below the patterns that are displayed using rectangles, the attribute transformations are given. Subsequently, a parallel combination and a sequential combination are applied and the effects are calculated. Showing that the expression "effort = 654 hours and requirements_document = verified" implies the goal completes the planning and demonstrates that a process combination has been found that promises to be a good basis for a project plan (based on the empirical evidence captured in the attribute transformations of the process patterns). For simplicity reasons, the attribute transformations in the example are described on the instance level (i.e., effort is given in absolute values). Typically, an additional mapping from the process pattern level to the instance level is necessary. For example, the effort needed for conducting a process pattern could be given as a percentage value of the overall effort, which is estimated during project planning. This effort needed for conducting a pattern may also be determined by a function on a value of the project characterization vector.

## 4. Evaluation

The method has been applied in a case study that focused on the planning of a large development effort for an embedded automation system. 42 process patterns were defined and applied. The goal of the evaluation was to qualitatively show the applicability of the method and to get feedback on how to improve the method. For the project two alternative development tools (Rapsody and StP) were considered that influenced many development activities. Therefore, two clusters of process pattern combinations were built. In the first cluster, 32 process pattern combinations needed to be assessed with respect to their suitability to fulfill the goal. In the second cluster, 15 alternatives needed to be assessed. In each cluster, less than 5 verification steps were necessary. Several combinations have been proven to be suitable. It was possible to choose among the results depending on priorities (e.g., minimize planned effort, optimize defect density rate). The project was conducted according to the plan using more than 20 students. The monitoring of the project data showed high plan adherence (based on qualitative judgement).

In addition, step 3 of the composition method has been evaluated in several case studies. This is described in detail in [10].

## 5. Conclusions and Future Research Topics

This paper sketched a method for composing process patterns in a goal-oriented way and showed initial evaluation results. The application of the method has shown that it is possible to provide mechanisms for systematically applying and using process patterns during project planning. Moreover, the method provides mechanisms for verifying that a certain combination of process patterns fulfills a project goal in a specific context. It should be mentioned that as with all models, the quality of the results heavily depends on the validity (i.e., mainly the empirical basis) of the models. Challenges for future research in this area are, for instance,

- the development of a graphical representation for process pattern combinations (especially mechanisms to visualize variability),
- automated support for verifying process pattern combinations against goals,
- gaining experience about appropriate granularity levels for process descriptions,
- generating process documentations from process pattern combinations,
- integrating patterns for software development activities and patterns/processes for system development activities (e.g., development activities in mechanical engineering).

## Acknowledgement

The author would like to thank Jens Heidrich from the Fraunhofer Institute for Experimental Software Engineering (IESE) who contributed to the implementation of the method, and Sonnhild Namingha from the Fraunhofer Institute for Experimental Software Engineering (IESE) for reviewing the first version of the article.

## References

[1] Linda C. Alexander and Alan M. Davis. Criteria for Selecting Software Process Models. In Proceedings of the 15th Annual International Computer Software & Applications Conference, pp. 521-528. IEEE Computer Society Press, Tokyo, 1991.

[2] Ralf Bergmann, Hector Muñoz-Avila, Manuela Veloso and E. Melis. Case-based Reasoning applied to Planning. In M. Lenz, B. Bartsch-Spörl, H.D. Burkhard

and S. Wess, Eds., Case-Based Reasoning Technology: From Foundations to Applications. Lecture Notes in Artificial Intelligence 1400, Springer Verlag, 1998.

[3] James O. Coplien. A Development Process Generative Pattern Language, chapter 13, pages 183--237. Addison-Wesley, Reading, MA, 1995.

[4] Khaled El Emam, Nazim H. Mahavji and K. Toubache. Empirically Driven Improvement of Generic Process Models. In Proceedings of the 8th International Software Process Workshop, 1993.

[5] Erich Gamma, Richard Helm, Ralph Johnson, John Vlissides. Design Patterns – Elements of Reusable Object-Oriented Software. Addison Wesley, 1995.

[6] Soeli T. Fiorini, Julio Cesar Sampaio do Prado Leite and Carlos José Pereira de Lucena. Reusing Process Patterns. 2nd International Workshop on Learning Software Organizations (LSO), 2000.

[7] Mariele Hagen and Volker Gruhn, Process Patterns – a Means to Describe Processes in a Flexible Way, Proceedings of the 5th International Workshop on Software Process Simulation and Modeling (ProSim 2004), Edinburgh, Scotland, United Kingdom, pp.50-56, May 24-25, 2004.

[8] Kathrin Lappe et al., Requirements Engineering Patterns – An Approach to Capturing and Exchanging RE Experience. Final Report from the WGREP, DESY 2004, Hamburg, Germany.

[9] M. Lipshutz, R. Creps und M. Simos. Organizational domain modeling (ODM) tutorial, January 1997.

[10] Jürgen Münch. Muster-Basierte Erstellung von Software-Projektplänen (in German). PhD Theses in Experimental Software Engineering, Vol. 10, ISBN: 3-8167-6207-7, Fraunhofer IRB Verlag, 2002.

[11] Jürgen Münch. Transformation-based Creation of Custom-tailored Software Process Models. Proceedings of the 5th International Workshop on Software Process Simulation and Modeling (ProSim 2004), Edinburgh, Scotland, United Kingdom, pp.50-56, May 24-25, 2004.

[12] Rubén Prieto-Díaz und Peter Freeman. Classifying Software for Reusability. IEEE Software, 4(1): 6-16, IEEE Computer Society, January 1987.

[13] Markus Schmitz, Jürgen Münch und Martin Verlage. Tailoring of large process models on the basis of MVP-L (in German). In Günther Müller-Luschnat, Sergio Montenegro, Ralf Kneuper, eds., Proceedings of the 4th workshops of the section 5.1.1 (GI), Berlin, March 1997